# Investigations of RhB18 Langmuir monolayer by Fluorescence Imaging Microscopy

S. A. Hussain, S. Chakraborty and D. Bhattacharjee*
Department of Physics, Tripura University, Suryamaninagar-799130

Email: tuphysic@sancharnet.in

**Abstract:** This communication reports the surface pressure vs. area per molecule isotherm and Fluorescence Imaging Microscopic studies of the formation of domain structure in the mixed Langmuir monolayer of RhB and Stearic acid (SA) at the air water interface. Strong repulsive interaction between the unlike components leads to the phase separation and formation of microcrystalline domains at the air water interface of the Langmuir monolayer. These domain can be directly visualized using fluorescence imaging microscope.

**Key words:** Isotherms, Langmuir films, Fluorescence Imaging Microscopy,

## 1. Introduction

Investigation of Langmuir monolayer on the air-water interface has been proved to be a relevant and fruitful activity for researchers in diverse fields of science [1-3]. Langmuir monolayer presents an opportunity to study the dependence of two-dimensional phase behaviour on molecular interactions and packing [4-7.]. The complexity of the phase and domain behaviours, both static and dynamic, makes these monolayers exhibit many examples of two dimensional condensed matter physics phenomena. A wide variety of molecules can be studied as monolayers, including "simple" single alkyl chain surfactants, phospholipids, organometallics, polymers and proteins.

The design and fabrication of molecular electronic devices using Langmuir-Blodgett technology encourages the seeking of a better characterization of the initial monolayer on the water surface.

Conventional techniques for studying Langmuir monolayer include surface pressure vs area per molecule isotherm study, surface potential and surface viscosity measurement and more recently developed methods such as in-situ FTIR spectroscopic studies as well as for domain structure visualization, in-situ Fluorescence Imaging Microscope (FIM) and Brewster Angle Microscope (BAM). [8,9.]

When transferred onto solid supports to form mono- and multilayered Langmuir-Blodgett Films, they have their potential use in the construction of molecular electronic devices. Most of the envisioned applications require incorporation of organic molecules having interesting chromophores. Studies on non-linear optical properties ferroelectricity, pyroelectric properties , photochromic effect are going on using LB films [10-14]

In this communication we have investigated the mixing behaviour of a laser dye Octadecyl Rhodamine B (RhB) and stearic acid in the mixed Langmuir monolayer on the water surface. Surface Pressure vs. area per molecule isotherm and fluorescence imaging microscopy (FIM) studies were used to characterize the monolayer films.

## 2.Experimental:

Octadecyl Rhodamin B (RhB, 99%,Molecular Probe) and SA (purity>99%) purchased from Sigma Chemical Company were used as received. Spectroscopy grade chloroform (SRL, India) was used as solvent and purity of the solvent was checked by fluorescence spectroscopy before use. Langmuir Blodgett (LB) film deposition instrument (Apex-2000C, India)was used for the study of isotherm characteristic as well also for the fabrication of Langmuir monolayer on the water surface of the LB trough by changing various parameters for fluorescence imaging microscopic study.

---
* Corresponding author

Phase contrast Fluorescence Imaging Microscope (FIM) (Model: Motic AE31) is attached with the LB instrument for in-situ study of the monolayer images.

For the study of isotherm characteristics as well as also for the fabrication of Langmuir monolayer, triple distilled deionised millipore water was used as subphase and the temperature was maintained at $24^0C$ with pH of the subphase at 6.5 in equilibrium with atmospheric $CO_2$. Solutions of octadecyl Rhodamine B (RhB), SA and RhB-SA mixture prepared in chloroform solvent and were spread on the water surface.

After a delay of 15 minutes to evaporate the solvent, the film at the air water interface was compressed slowly at the rate of $2x10^{-3}nm^2mol^{-1}s^{-1}$ to study the surface pressure vs. area per molecule (π-A) isotherm. FIM images of the monolayers of pure components (RhB and SA) were taken at 25 mN/m surface pressure. FIM images of mixed monolayer were taken at various surface pressures starting from 0 mN/m upto 25 mN/m.;

## 3 Results and discussions
### 3.1. Surface pressure vs. area per molecule Isotherm

Fig.1 shows the surface pressure vs. area per molecule isotherms of pure SA, pure RhB and RhB-SA mixed film having 0.1 mole fractions of RhB. The isotherm of pure SA is consistent with the works mentioned in the literature [15]. The area per molecules are 0.23 and 0.21 for pure SA at the surface pressures 15 and 25 mN/m and collapse pressures at about 55mN/m. Isotherm of pure RhB shows an infliction point of about 32 mN/m with collapse pressure at about 40 mN/m. This inflection point indicates the rearranging and orientation of RhB molecules at the air water interface. Isotherm of mixed monolayer of RhB–SA at 0.1 molefraction of RhB lies in between the isotherm of pure SA and RhB. It has a collapse pressure at 30 mN/m. This low value of collapse pressure is an indication of strong repulsive interaction between the unlike components which may result in the formation of aggregate of like components resulting in the formation of distinct domain of RhB and SA.

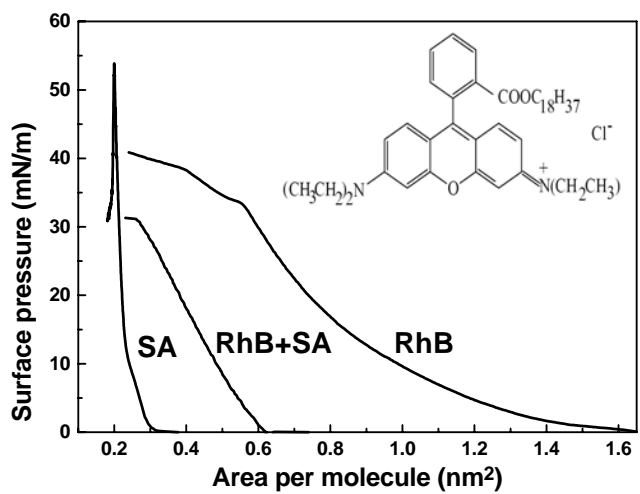

Fig. 1 Surface pressure area per molecule isotherm of SA, RhB and RhB-SA (0.1 : 0.9 molar ratio) mixture. Inset: molecular structure of RhB.

## 3.2 In-situ Fluorescence Imaging Microscopic (FIM) study of Langmuir monolayer

Fig. 2a and 2b show the FIM images of pure SA monolayer and Pure RhB monolayer at the air water interface at a surface pressure of 25 mN/m. Due to absence of any fluorescence , SA monolayer gives a dark image which is uniform throughout the film, whereas RhB monolayer fluorescence image gives a bright crimson red illumination. This illumination is uniform throughout the film indicating formation of uniform monolayer of RhB.

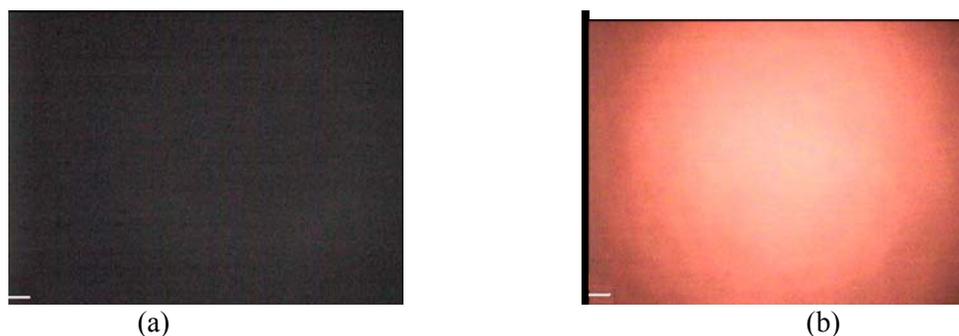

(a)　　　　　　　　　　　　　　　　　　(b)

Fig. 2 Fluorescence microscopic images of (a) SA and (b) RhB monolayers respectively at the air-water interface. The scale bar in the lower left represents 10 µm.

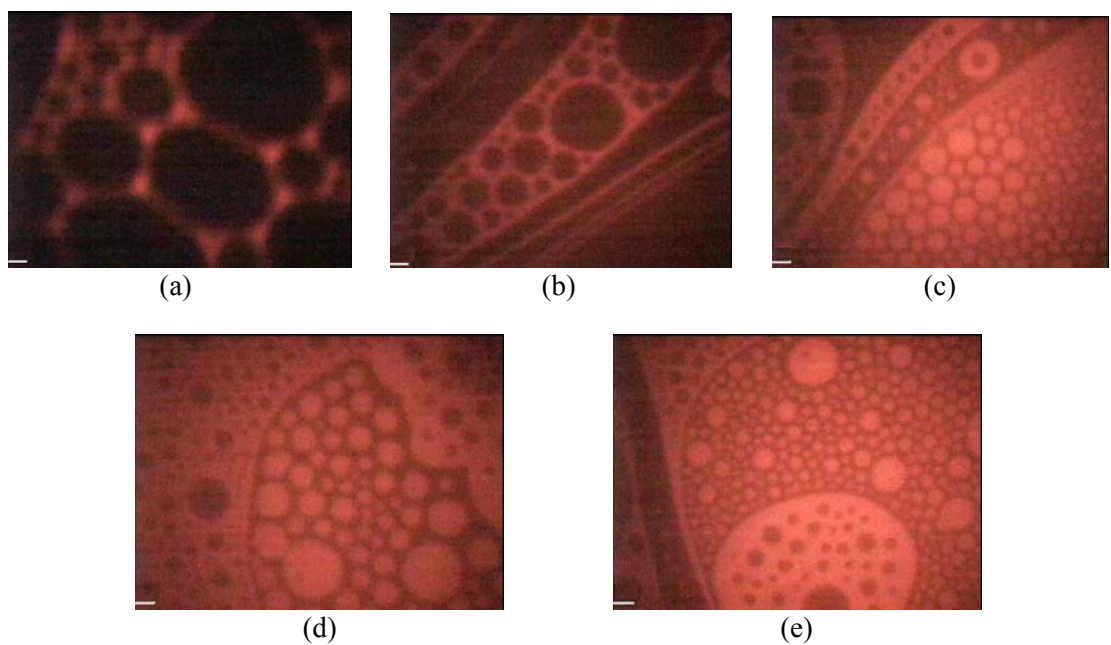

(a)　　　　　　　　(b)　　　　　　　　(c)

(d)　　　　　　　　(e)

Fig: 3 Fluorescence Imaging Microscopes of Langmuir monolayer of RhB mixed with SA on water subphase at molar ratio of RhB:SA=0.1:0.9 at different pressures viz (a) 0 mN/m, (b) 5 mN/m, (c) 15 mN/m, (d) 20 mN/m and (e) 25 mN/m. The scale bar in the lower left represents 10 µm

When mixed monolayer is formed on the air water interface then due to strong repulsive interactions between unlike components phase separation occurs leading to the formation of aggregates and domain structures of micrometer range. Figures 3a to 3e show the fluorescence microscopic images of mixed Langmuir monolayer taken at various surface pressures starting from 0 mN/m up to 25 mN/m. It is interesting to note that in the mixed film, phase separation is started even at 0 mN/m surface pressure (Fig. 3a). At 10 mN/m surface pressure (Fig. 3b) distinct circular shaped black patches are observed having bordered with crimson red colour. The circular black patches are due to SA where as crimson red coloured bordering areas are due to RhB. At higher surface pressures of 15 mN/m upto 25 mN/m the contrast become opposite. i.e., distinct crimson coloured circular patches with lightly illuminated background is observed. This brightly illuminated circular patches are due to the formation of distinct domains of RhB microcrystalline aggregates. The dimension of these domains ranges typically from about 5 µm to 25 µm. The lightly illuminated background clearly indicates that RhB domains are pushed up on the SA film and covers the SA monolayer.

## 4. Conclusion

In conclusion our results show that due to strong repulsive interactions due to unlike components in the mixed monolayers of RhB-SA, phase separation occurs. This leads to the formation of distinct circular shaped domain structures of RhB micro-crystals. FIM images at high surface pressure, the dimension of the domain ranges between 5 µm to 25 µm.

## 5. Acknowledgement

The authors are grateful to DST, Govt. of India for providing financial support through the project no. DST Project No: SR/S2/LOP-19/07.